\documentclass[english]{article}
\usepackage[affil-it]{authblk}
\usepackage[T1]{fontenc}
\usepackage[latin9]{inputenc}
\usepackage{mathrsfs}
\usepackage{refstyle}
\usepackage{color,hyperref}
    \catcode`\_=11\relax
    \newcommand\email[1]{\_email #1\q_nil}
    \def\_email#1@#2\q_nil{%
      \href{mailto:#1@#2}{{\emailfont #1\emailampersat #2}}
    }
    \newcommand\emailfont{\sffamily}
    \newcommand\emailampersat{{\color{red}\small@}}
    \catcode`\_=8\relax

\makeatletter

\RS@ifundefined{subref}
  {\def\RSsubtxt{section~}\newref{sub}{name = \RSsubtxt}}
  {}
\RS@ifundefined{thmref}
  {\def\RSthmtxt{theorem~}\newref{thm}{name = \RSthmtxt}}
  
  {}
\RS@ifundefined{lemref}
  {\def\RSlemtxt{lemma~}\newref{lem}{name = \RSlemtxt}}
  {}

\usepackage{amsmath}

\makeatother

\usepackage{babel}
\begin{document}

\title{Galilean covariance of quantum-classical hybrid systems of the Sudarshan
type. }
\author{A. D. Berm\'udez Manjarres}
\affil{\footnotesize Universidad Distrital Francisco Jos\'e de Caldas\\ Cra 7 No. 40B-53, Bogot\'a, Colombia\\ \email{ad.bermudez168@uniandes.edu.co}}
\author{N. Mar\'in-Medina}
\affil{\footnotesize Laboratorio de Investigaci\'on en Electronica y Redes- LIDER, Facultad de Ingenier\'ia Universidad Distrital Francisco Jos\'e de Caldas\\ 110111, Bogot\'a, Colombia\\ \email{nmarin@uniandes.edu.co}}

\maketitle
\begin{abstract}
We revisit quantum-classical hybrid systems of the Sudarshan type
under the light of Galilean covariance. We show that these kind of
hybrids cannot be given as a unitary representation of the Galilei
group and at the same time conserve the total linear momentum unless
the interaction term only depends on the relative canonical velocities.
\end{abstract}

\section{introduction}

Symmetry considerations play a fundamental role in modern physics,
among other things, they serve as guiding principles for the development
of new theories. It has been argued that there is a possibility that
mesoscopic systems could be better described by a quantum--classical
hybrid theory, whether it is a known theory or a yet-to-be discovered
hybrid \cite{barcelo}. Such theory might not be derivable as any
limit of pure quantum mechanics. Given that both non-relativistic
classical and quantum mechanics are Galilei covariant, it seems natural
to impose Galilei covariance to any hybrid quantum-classical theory.
In this work we investigate the consequences of Galilei covariance
on an specific kind of hybrid systems, the sometimes called Sudarshan
hybrids \cite{sudarshan0,sudarshan1,sudarshan2,sudarshan3}. 

Motivated by the measurement problem of quantum mechanics, Sudarshan
and coworkers used the operational version of classical mechanics
called the Koopman-von Neumann theory \cite{KvN1,KvN2} (hereafter
abbreviated as the KvN theory) to couple a classical system with a
quantum one. The KvN theory has the advantage of expressing classical
mechanics in the same mathematical formalism used for quantum mechanics,
i.e., the mathematics of operators acting on a Hilbert space. Sudarshan's
hybrid theory was initially aimed to study the interaction of a quantum
system with a classical measuring object. Peres and Terno objected
the consistency of the hybrids of the Sudarshan type \cite{peres,peres2}.
Objections to general hybrid systems in more general contexts have
been also stated\cite{salcedo1,salcedo2}. The above criticisms presuppose
that the hybrid systems are a kind of partial classical limit of a
fully quantum situation. Barcel\'{o} \emph{et al }\cite{barcelo}\emph{
}take the raised objections as an opportunity for the development
of new theories in the mesoscopic level, theories that are not derivable
from pure quantum mechanics.

It is natural to posit that any new theory in the non-relativistic
regime should be Galilei covariant. The covariance of physical theories
under groups of space-time transformation is the subject of too many
publications to be listed here, we will only cite a handful of them.
Non-relativistic quantum mechanics can be obtained as a irreducible
unitary representation of the Galilei group \cite{quantum1,quantum2,jordan,ballentine}
while the structure of relativistic quantum field theories are determined
by the properties of the Poincare group \cite{weinberg}. On the classical
part, non-relativistic classical mechanics has been derived from the
structure of the Galilei group in the context of Lagrangian mechanics
\cite{quantum1,landau} and as a canonical representation in terms
of Poisson brackets in Hamiltonian mechanics \cite{canonical,canonical2}. 

The differences in the approaches between the classical and the quantum
representations of the Galilei group (unitary vs canonical representations)
are due to the different mathematical formalism in which each theory
is usually expressed. Nevertheless, the KvN theory and, hence, classical
mechanics can also be obtained by considering an irreducible unitary
representation of the Galilei group \cite{KvN3}. Thus, the Galilei
covariance of the quantum and classical sectors of the Sudarshan Hybrids
can both be studied within the context of unitary representations
of Lie groups. The quantum and classical unitary representations of
the Galilei group differ in several critical aspects that will be
reviewed later in the main part of this work.

The main purpose of this work is to study the restrictions imposed
by Galilei covariance on the hybrids of the Sudarshan type, whether
they are understood as a partial limit of a quantum theory or a fully
new theory on their own, .

This work is organized as follows. In section 2, a review of the KvN
formalism is given; only the most necessary concepts are presented
but references are given for a more in depth treatment of the theory. 

In section 3, we present a review of the quantum and the less-known
classical unitary representation of the Galilei algebra. Again, only
the necessary tools required for the study of Sudarshan hybrids are
presented. For both cases only spinless particles are considered.

In section 4, the Galilei algebra for the hybrid system is constructed
and the restriction imposed on the interaction Hamiltonian are studied.
The main results of this work are derived here. The results can be
summarized as follows:
\begin{enumerate}
\item Under very natural assumptions, the Galilei covariance of Sudarshan
hybrids restrict the choices of interaction terms between the quantum
and classical systems. A list of all possible interaction terms is
given.
\item None but one of the possible interaction terms conserve the total
(quantum + classical) linear momentum. 
\item Conservation of total momentum prevents any quantum back reaction
on the classical observable variables.
\end{enumerate}
The Einstein summation convention is used through all this work. 

\section{The KvN formalism}

In the KvN operational formulation of classical mechanics of a point
particle the position and momentum $(\hat{\mathbf{q}},\hat{\mathbf{p}})$
are understood as Hermitian operators that commute with each other,
and, therefore, no uncertainty principle is present between them.
The theory also introduces two auxiliary Hermitian vector operators
$\hat{\lambda}_{\mathbf{q}}$ and $\hat{\lambda}_{\mathbf{p}}$ that
are understood to have no direct physical meaning and are deemed as
unobservables. These four operators obey the following commutation
relations

\begin{align}
\left[\hat{q}_{i},\hat{q}_{j}\right] & =\left[\hat{p}_{i},\hat{p}_{j}\right]=\left[\hat{q}_{i},\hat{p}_{j}\right]=\left[\hat{q}_{i},\hat{\lambda}_{p_{j}}\right]=\left[\hat{p}_{i},\hat{\lambda}{}_{q_{j}}\right]=0,\\
\left[\hat{q}_{i},\hat{\lambda}{}_{q_{j}}\right] & =\left[\hat{p}_{i},\hat{\lambda}_{p_{j}}\right]=i\delta_{ij}.
\end{align}

The operators of the classical system act on a Hilbert space $\mathcal{H}_{cl}$
whose elements are vectors of the form $\left|\psi\right\rangle =\int\left\langle \mathbf{q},\mathbf{p}\right|\left.\psi\right\rangle \left|\mathbf{q},\mathbf{p}\right\rangle \,d\mathbf{q}d\mathbf{p}$
such that the complex wavefunction $\psi(\mathbf{q},\mathbf{p})=\left\langle \mathbf{q},\mathbf{p}\right.\left|\psi\right\rangle $
is square integrable. The wavefunction $\psi(\mathbf{q},\mathbf{p})$
represents the probability amplitude of finding a particle in certain
region of phase space, while the quantity $\rho=\left|\psi(\mathbf{q},\mathbf{p})\right|^{2}$
is the probability density in phase space used in classical statistical
mechanics. The kets $\left|\mathbf{q},\mathbf{p}\right\rangle $ are
eigenfunctions of $\hat{\mathbf{q}}$ and $\hat{\mathbf{p}}$ 

\begin{align}
\hat{q}_{i}\left|\mathbf{q},\mathbf{p}\right\rangle  & =q_{i}\left|\mathbf{q},\mathbf{p}\right\rangle ,\nonumber \\
\hat{p}_{i}\left|\mathbf{q},\mathbf{p}\right\rangle  & =p_{i}\left|\mathbf{q},\mathbf{p}\right\rangle ,
\end{align}
and they obey the orthonnormality condition $\left\langle \mathbf{q}',\mathbf{p}'\right.\left|\mathbf{q},\mathbf{p}\right\rangle =\delta(\mathbf{q}-\mathbf{q}')\delta(\mathbf{p}-\mathbf{p}').$ 

If the classical system is governed by a Hamiltonian function $H_{c}(\mathbf{q},\mathbf{p})$,
then the evolution of $\left|\psi\right\rangle $ is ruled by a Schr\"odinger
equation

\begin{equation}
\frac{d}{dt}\left|\psi(t)\right\rangle =-i\hat{H}_{cl}\left|\psi(t)\right\rangle ,\label{scrho}
\end{equation}
where $\hat{H}_{cl}$, known as the Liouvillian operator, is given
by

\begin{equation}
\hat{H}_{cl}=\nabla_{\mathbf{p}}H_{c}\cdot\hat{\lambda}_{\mathbf{q}}-\nabla_{\mathbf{q}}H_{c}\cdot\hat{\lambda}_{\mathbf{p}}.\label{Hc-1}
\end{equation}
For the typical classical Hamiltonian of a single particle $H_{c}=\frac{\mathbf{p}^{2}}{2m}+V(q)$\footnote{For simplicity, and since the main interest is in the classical-quantum
coupling, we do not consider here the case where the classical system
interact with an external force that have to be handled with the introduction
of a vector potential.}, the Liouvillian operator takes the form

\begin{equation}
\hat{H}_{cl}=\frac{1}{m}\hat{\mathbf{p}}\cdot\hat{\lambda}_{\mathbf{q}}-\nabla_{\mathbf{q}}V(q)\cdot\hat{\lambda}_{\mathbf{p}}.
\end{equation}

In the ``wave mechanics'' version of the theory, $\hat{\mathbf{q}}$
and $\hat{\mathbf{p}}$ act as multiplicative operators on $\psi(\mathbf{q},\mathbf{p})$,
while $\hat{\lambda}_{\mathbf{q}}$ and $\hat{\lambda}_{\mathbf{p}}$
act as derivatives

\begin{equation}
\hat{\lambda}_{\mathbf{q}}=-i\nabla_{\mathbf{q}};\;\hat{\lambda}_{\mathbf{p}}=-i\nabla_{\mathbf{p}}.\label{Id}
\end{equation}
In this version of the theory the Liouvillian becomes a differential
operator whose action on phase space functions $\psi(\mathbf{q},\mathbf{p})$
is given by

\begin{equation}
\hat{H}_{cl}\psi=-i\left\{ \psi,H_{c}\right\} ,\label{PB}
\end{equation}
where the above bracket is the Poisson bracket of Hamiltonian mechanics.

The Schr\"odinger equation (\ref{scrho}) then reduces to 

\begin{equation}
\frac{\partial\psi}{\partial t}=-\left\{ \psi,H_{c}\right\} ,
\end{equation}
 Since $\hat{H}_{cl}$ is linear in $\hat{\lambda}_{\mathbf{q}}$
and $\hat{\lambda}_{\mathbf{p}}$, $\rho$ obeys the same equation
as $\psi$

\begin{equation}
\frac{\partial\rho}{\partial t}=-\left\{ \rho,H_{c}\right\} .\label{L}
\end{equation}
Eq.(\ref{L}) is the Liouville equation of classical statistical mechanics,
and it is the proof that the abstract formulation in terms of operators
of the KvN theory is equivalent to Hamiltonian mechanics.

Finally, let us mention that the operators of the set $\left\{ \hat{\mathbf{q}},\hat{\mathbf{p}},\hat{\lambda}_{\mathbf{q}},\hat{\lambda}_{\mathbf{p}}\right\} $
are self-andjoint on $\mathcal{H}_{cl}$, this set is also irreducible
in this Hilbert space\footnote{There is a possibility of adding internal degrees of freedom to the
theory, $\left|\mathbf{q},\mathbf{p},\sigma\right\rangle $, so $\left\{ \hat{\mathbf{q}},\hat{\mathbf{p}},\hat{\lambda}_{\mathbf{q}},\hat{\lambda}_{\mathbf{p}}\right\} $
is not irreducible on $\mathcal{H}_{cl}$. In particular $\sigma$
may designate a classical spin for the classical point particle. The
Galilei algebra allows for such classical spin, but here we only consider
spinless particles.}. 

\section{Review of the Quantum and the Classical representations of the Galilei
Algebra }

The proper Galilei group is a ten parameter Lie group that consists
of space-time translations, rotations and transformations to moving
frames (boosts). The group elements are to be realized by unitary
transformation acting on the appropriated Hilbert space. To each space-time
transformation there is associated an Hermitian operator, the generator
of the transformation. The derivation of the Lie algebra associated
to the Galilei group can be found, for example, in \cite{jordan}. 

While both quantum and classical mechanics are associated to the same
brackets relations, the realization of the Lie algebra generators
is different in each case. In particular, the classical unitary representation
of the Galilei algebra has only been recently considered \cite{KvN3}
and it is less straightforward than the quantum representation (since
the classical case necessarily involves non-observable operators).
For the above reasons, both representations shall be reviewed separately.
Even if the reader is well versed in the Galilei group and algebra,
it is recommended to follow closely the next two subsections as they
fix the notation used later. 

\subsection{Quantum Representation}

The position and momentum operator for the quantum system will be
designated by $\hat{\mathbf{r}}$ and $\hat{\mathbf{k}}$, respectively.
They, of course, obey the Heisenberg commutation relation\footnote{Through all this work we set $\hbar=1$}
$\text{\ensuremath{\left[\hat{r}_{i},\hat{k}_{j}\right]}}=i\delta_{ij}$.
The generators of the Galilei algebra for the quantum sectors are:
the translation operator, in the quantum case it coincides whit the
momentum $\hat{\mathbf{k}}$; an operator for rotations $\hat{\mathbf{j}}$;
the operator for the Galilean boosts $\hat{\mathbf{g}}$; the Hamiltonian
$\hat{H}_{\mathrm{Q}}$ is the time translation operator; and the
central charge of the algebra, $M$ (which will be considered here
just as a real number). The Galilei algebra is a set of commutator
equations that involves the generators of the space-time transformation,
these equations are

\begin{align}
\left[\hat{k}_{i},\hat{k}_{j}\right] & =\left[\hat{g}_{i},\hat{g}_{j}\right]=\left[\hat{j}_{i},\hat{H}_{\mathrm{Q}}\right]=\left[\hat{k}_{i},\hat{H}_{\mathrm{Q}}\right]=0,\nonumber \\
\left[\hat{j}_{i},\hat{j}_{j}\right] & =i\varepsilon_{ijk}\hat{j}_{i};\;\left[j_{i},\hat{k}_{j}\right]=i\varepsilon_{ijk}\hat{k}_{k};\nonumber \\
\left[\hat{j}_{i},\hat{g}_{j}\right] & =i\varepsilon_{ijk}\hat{g}_{k};\;\left[\hat{k}_{i},\hat{g}_{j}\right]=i\delta_{ij}M;\:\left[\hat{g}_{i},\hat{H}_{\mathrm{Q}}\right]=i\hat{k}_{i}.\label{galileialgebra}
\end{align}
In the quantum case the generators of the algebra have well established
physical meaning. $M$ is the mass (and for a physical representation
$M$ is strictly a positive number), $\hat{\mathbf{k}}$ is the momentum,
$\hat{\mathbf{j}}$ is the angular momentum, $\hat{H}_{\mathrm{Q}}$
is the energy (of a free particle) and $\hat{\mathbf{g}}$ is a physical
quantity sometimes called the dynamic mass moment. 

For a single spinless particle the momentum and position form an irreducible
set of operators. The generators of the algebra are given in terms
of $\hat{\mathbf{r}}$ and $\hat{\mathbf{k}}$ by

\begin{align}
\hat{\mathbf{j}} & =\hat{\mathbf{r}}\times\hat{\mathbf{k}},\\
\hat{\mathbf{g}} & =M\hat{\mathbf{r}}-t\hat{\mathbf{k}},\\
\hat{H}_{\mathrm{Q}} & =\frac{\hat{k}^{2}}{2M}.\label{HQ}
\end{align}
For systems of several particles the generators for individual particles
can be combined to give generators for the entire systems. For example,
for two particles the space translation are generated by the total
linear momentum $\hat{\mathbf{k}}=\hat{\mathbf{k}}_{1}+\hat{\mathbf{k}}_{2}$
while rotations are generated by the total angular momentum $\hat{\mathbf{j}}=\hat{\mathbf{j}}_{1}+\hat{\mathbf{j}}_{2}$.
The total Hamiltonian is allowed to have a extra term that account
for the interaction between the two particles.

\begin{equation}
\hat{H}_{\mathrm{Q}}=\frac{\hat{k}_{1}^{2}}{2M_{1}}+\frac{\hat{k}_{2}^{2}}{2M_{2}}+V.
\end{equation}
The Galilei algebra commutations relations not involving the Hamiltonian
are all identically satisfied. The remaining relations give restrictions
to the potential energy $V$. It can be shown that, in order for all
the commutation equations to be satisfied, $V$ can only depend on
scalar combinations of the relative position $\hat{\mathbf{r}}_{1}-\hat{\mathbf{r}}_{2}$
and the relative momentum $\hat{\mathbf{k}}_{1}-\hat{\mathbf{k}}_{2}$
\cite{jordan}. It follows that the allowed total Hamiltonian is such
that the total linear and angular momentum are constant of the motion.

\subsection{Classical Representation}

The KvN version of classical mechanics is most usually obtained starting
from the Liouville equation, but it can also be derived as an irreducible
unitary representation of the Galilei group where the operators of
the algebra act on the Hilbert space $\mathcal{H}_{cl}$. To avoid
confusion, when dealing with the classical representation we make
the following change of notation for the operators of the Galilei
algebra: $\mathcal{\hat{J}}$ is the operator of rotations, $\hat{\mathcal{G}}$
is associated to Galilean boosts, the operator $\hat{H}_{cl}$ gives
time translation and $\hat{\lambda}_{\mathbf{q}}$ gives space translations.
We maintain the symbol $M$ for the central charge appearing in the
commutation relation for $\hat{\lambda}_{\mathbf{q}}$ and $\hat{\mathcal{G}}$

\begin{equation}
\left[\hat{\lambda}_{q_{i}},\hat{\mathcal{G}}_{j}\right]=i\delta_{ij}M.
\end{equation}

In term of the irreducible set $\left\{ \hat{\mathbf{q}},\hat{\mathbf{p}},\hat{\lambda}_{\mathbf{q}},\hat{\lambda}_{\mathbf{p}}\right\} $
the elements of the Galilei algebra are given by 

\begin{eqnarray}
\mathcal{\hat{J}}_{i} & = & \varepsilon_{ijk}\left(\hat{q}_{j}\hat{\lambda}_{q_{k}}+\hat{p}_{j}\hat{\lambda}_{p_{k}}\right),\nonumber \\
\hat{\mathcal{G}}_{i} & = & -\hat{\lambda}_{q_{i}}t-m\hat{\lambda}_{p_{i}},\nonumber \\
\hat{H}_{cl} & = & \frac{1}{m}\hat{p}_{i}\hat{\lambda}_{q_{i}},\label{cla rep}
\end{eqnarray}
where $m$ is a positive number interpreted as the mass. Let us stress
here that the mass $m$ should not be confused with the central charge
$M$. It can be checked by direct computation that operators (\ref{cla rep})
obey, with one notable exception, the same commutation relations as
their quantum equivalents with the following replacement

\begin{align}
\hat{k}_{i} & \rightarrow\hat{\lambda}_{q_{i}},\nonumber \\
\hat{j}_{i} & \rightarrow\hat{\mathcal{J}}_{i},\nonumber \\
\hat{g}_{i} & \rightarrow\hat{\mathcal{G}}_{i},\nonumber \\
\hat{H}_{\mathrm{Q}} & \rightarrow\hat{H}_{cl}.
\end{align}

In the ``wave mechanics'' version of the KvN theory, the Eqs.(\ref{Id})
and (\ref{PB}) explain how to get the Liouvillian from the classical
Hamiltonian function via the Poisson bracket. The time translation
operator appearing in (\ref{cla rep}) is obtained when the Hamiltonian
of a free particle is used $H_{c}=\frac{\mathbf{p}^{2}}{2m}$. The
others generators can be written as Poisson brackets of phase space
functions as follows $\hat{\lambda}_{\mathbf{q}}=-i\left\{ \cdot,\mathbf{p}\right\} $,
$\mathcal{\hat{J}}=-i\left\{ \cdot,\mathbf{L}\right\} $ and $\hat{\mathcal{G}}=-i\left\{ \cdot,\mathbf{\mathbf{g}}\right\} $,
where $\mathbf{L}=\mathbf{r}\times\mathbf{p}$ is the angular momentum
and $\mathbf{g}=m\mathbf{q}-t\mathbf{p}$ is the dynamic mass moment.

The only difference between the quantum and classical representation,
albeit a remarkable one, comes from the central charge. In the quantum
case $M$ is to be a positive number in order to have a physical representation
of the group. However, in the classical case the only possible choice
is $M=0$, so the central charge cannot be interpreted as the mass.
In fact, the mass $m$ does not appear in the relations of in the
Galilei algebra at all. The only different commutation relation, then,
reads 

\begin{equation}
\left[\hat{\lambda}_{q_{i}},\hat{\mathcal{G}}_{j}\right]=0.\label{G}
\end{equation}
Eq. (\ref{G}) can be derived without any use of results from analytical
dynamics \cite{KvN3}, but it is also a direct consequence of the
way the KvN theory assigns operators to phase space functions. It
can be checked that the differential operators $\mathcal{\hat{J}}=-i\left\{ \cdot,\mathbf{L}\right\} $
and $\hat{\mathcal{G}}=-i\left\{ \cdot,\mathbf{\mathbf{g}}\right\} $
commute.

The above is highly striking when compared to the quantum case since
quantums representation with vanishing $M$ are unphysical\cite{quantum1}.
However, this situation is not a problem since, unlike the quantum
case, in the classical representation of the Galilei group the elements
of the algebra have no direct physical meaning. For example, $\hat{\mathcal{J}}$
is an operator of rotations but not an angular momentum while $\hat{H}_{cl}$
generates time translation but its spectrum is not related to the
energy of the system. By the same token, $M$ needs not to be related
with the mass and it can vanish without causing any problem. 

Just as in the quantum case, the Galilei algebra impose restrictions
on the interaction between classical particles. It can be shown that
the forces between the particles can only depend on scalar combinations
of the relative position and the relative velocities, the conservation
of the total linear momentum is closely related. The result just mentioned
is known from analytical mechanics\cite{canonical2} but it can also
be obtained from the classical unitary representation of the Galilei
group in the context of the KvN theory \cite{KvN3}. Let us take as
an example a system of only two particles with velocity independent
interaction. For this system the Liouvillian takes the form

\begin{equation}
\hat{H}_{cl}=\frac{1}{m_{1}}\hat{\mathbf{p}}_{1}\cdot\hat{\lambda}_{\mathbf{q_{1}}}+\frac{1}{m_{2}}\hat{\mathbf{p}}_{2}\cdot\hat{\lambda}_{\mathbf{q_{2}}}-\nabla_{\mathbf{q_{1}}}V\cdot\hat{\lambda}_{\mathbf{p}_{1}}-\nabla_{\mathbf{q}_{2}}V\cdot\hat{\lambda}_{\mathbf{p}_{2}}.\label{Htwo}
\end{equation}
The Eq.(\ref{Htwo}) can be obtained from symmetry principles though
the procedure is somewhat convoluted \cite{KvN3}. A second and simpler
alternative consist in using the Poisson bracket formula (\ref{PB})
with the classical Hamiltonian given by $H_{c}=\frac{\mathbf{p}_{1}^{2}}{2m}+\frac{\mathbf{p}_{1}^{2}}{2m}+V(\mathbf{q}_{1}-\mathbf{q}_{2})$
as follows

\begin{align}
\hat{H}_{cl} & =-i\left\{ \cdot,H_{c}\right\} =\frac{-i}{m_{1}}\mathbf{p}_{1}\cdot\nabla_{\mathbf{q_{1}}}+\frac{-i}{m_{2}}\mathbf{p}_{2}\cdot\nabla_{\mathbf{q_{2}}}\nonumber \\
 & +i\nabla_{\mathbf{q_{1}}}V\cdot\nabla_{\mathbf{p_{2}}}+i\nabla_{\mathbf{q}_{2}}V\cdot\nabla_{\mathbf{p_{2}}}.\label{PB2}
\end{align}

The Eq.(\ref{Htwo}) is recovered by identifying the auxiliary operators
$\hat{\lambda}$ in (\ref{PB2}) with the use of the formulas given
in (\ref{Id}). Since the potential depends on $\hat{\mathbf{q}}_{1}-\hat{\mathbf{q}}_{2}$,
it follows that $\nabla_{\mathbf{q_{1}}}V=-\nabla_{\mathbf{q}_{2}}V$.
The total linear momentum $\hat{\mathbf{p}}_{1}+\hat{\mathbf{p}}_{2}$
is a conserved quantity as it commutes with $\hat{H}_{cl}$. Let us
notice that, while related, the conservation of the total momentum
should not be confused with the conservation of the total translation
operator $\hat{\lambda}_{\mathbf{q}_{1}}+\hat{\lambda}_{\mathbf{q}_{2}}$.

\section{Sudarshan Hybrids}

Having a formulation of classical mechanics in terms of operators
acting on a Hilbert space, the quantum and classical sectors can be
treated on the same footing. The Sudarshan hybridization consist in
coupling the quantum and classical systems using a tensor product,
just as to pure quantum systems are coupled in the standard quantum
theory. The space of states for the hybrid system is $\mathcal{H}_{T}=\mathcal{H}_{\mathrm{Q}}\mathrm{\otimes}\mathcal{H}_{cl}$.
It follows from its definition that in the joint system the quantum
operators commute with the classical ones. 

Just as in pure quantum mechanics, the generators of space-time transformation
with purely geometrical interpretation (space translation, rotations
and boosts) acting on the joint space $\mathcal{H}_{T}$ are made
by adding the corresponding operators for the individual systems

\begin{align}
\hat{\mathscr{P}}_{i} & =\hat{k}_{i}+\hat{\lambda}_{q_{i}},\nonumber \\
\hat{G}_{i} & =\hat{g}_{i}+\hat{\mathcal{G}}_{i},\nonumber \\
\hat{\mathscr{J}}_{i} & =\hat{j}_{i}+\hat{\mathcal{J}}_{i}.\label{totals}
\end{align}
It is worth noting that while $\hat{\mathscr{P}}$ and $\hat{\mathscr{J}}$
generate translations and rotations for the joint system, they are
not the total linear momentum nor the total angular momentum. When
the systems are non-interacting, the total momentum is given by $\hat{\mathbf{P}}_{t}=\hat{\mathbf{k}}+\hat{\mathbf{p}}$,
and it is a constant of the motion. It seems most natural to keep
$\hat{\mathbf{P}}_{t}$ as the total momentum when the interaction
is switched on.

The generator of time translation is allowed to have an extra term
that accounts for the interaction of the two systems

\begin{equation}
\hat{H}_{T}=\hat{H}_{\mathrm{Q}}+\hat{H}_{cl}+\hat{H}_{int}.\label{HT}
\end{equation}
All commutators of the Galilei algebra not involving $\hat{H}_{T}$
are identically satisfied. In particular, the mass of the quantum
system (and not the total mass) becomes again the central charge of
the coupled algebra

\begin{equation}
\left[\hat{\mathscr{P}}_{i},\hat{G}_{i}\right]=\delta_{ij}M.
\end{equation}
On the other hand, the commutators involving $\hat{H}_{T}$ restrict
the possible choices for $\hat{H}_{int}$. From the commutation relations

\[
\left[\hat{\mathscr{J}}_{i},\hat{H}_{\mathrm{T}}\right]=\left[\hat{\mathscr{P}}_{i},\hat{H}_{\mathrm{T}}\right]=0;\:\left[\hat{G}_{i},\hat{H}_{T}\right]=i\hat{\mathscr{P}}_{i},
\]
it follows that the interaction term has to obey the following conditions

\begin{subequations}

\begin{align}
\left[\hat{\mathscr{P}}_{i},\hat{H}_{int}\right] & =0,\label{Ha}\\
\left[\hat{G}_{i},\hat{H}_{int}\right] & =0,\label{Hb}\\
\left[\hat{\mathscr{J}}_{i},\hat{H}_{int}\right] & =0.\label{Hc}
\end{align}

\end{subequations}

Before studying the restriction imposed on $\hat{H}_{int}$ by the
above equations, let us discuss a further requirements that comes
solely from the classical sector. From the expression given for $H_{cl}$
of a free particle (\ref{cla rep}) and the equation $\hat{\mathbf{p}}=im\left[\hat{\mathbf{q}},\hat{H}_{T}\right]$,
it is obtained that $\left[\hat{\mathbf{q}},\hat{H}_{int}\right]=0$.
The above means the interaction term is independent of $\hat{\lambda}_{\mathbf{q}}$.
On the other hand, the acceleration operator for the classical sector
is related to the classical momentum by

\begin{equation}
\hat{\mathbf{a}}=\frac{i}{m}\left[\hat{\mathbf{p}},\hat{H}_{T}\right]=\frac{i}{m}\left[\hat{\mathbf{p}},\hat{H}_{int}\right].\label{acc}
\end{equation}
If the acceleration is to be an observable, then it cannot depend
on the non-observable operators $\hat{\lambda}_{\mathbf{q}}$ and
$\hat{\lambda}_{\mathbf{p}}$. Hence, from (\ref{acc}), we find again
that $\hat{H}_{int}$ has to be independent on $\hat{\lambda}_{\mathbf{q}}$,
and, additionally, $\hat{H}_{int}$ has to be at most linear on $\hat{\lambda}_{\mathbf{p}}$.

Now, returning to the conditions from the Galilei algebra, we have
from Eq.(\ref{Ha}) that $\hat{H}_{int}$ can only be function of
the relative position $\hat{\mathbf{r}}-\hat{\mathbf{q}}$ since other
combinations of $\hat{\mathbf{r}}$ and $\hat{\mathbf{q}}$ do not
commute with $\hat{\mathscr{P}}_{i}$. By the same token, Eq.(\ref{Hb})
implies that $\hat{H}_{int}$ can only be function of the relative
\emph{velocity} $\frac{\hat{\mathbf{k}}}{M}-\frac{\hat{\mathbf{p}}}{m}$.
No equation from the algebra restrict the dependence of $\hat{H}_{int}$
on $\hat{\lambda}_{\mathbf{p}}$. Finally, Eq.(\ref{Hc}) can only
be satisfied if $\hat{H}_{int}$ is an scalar operator. Hence, the
interaction term can only be constructed from the following scalar
combinations: $\left(\hat{\mathbf{r}}-\hat{\mathbf{q}}\right)^{2}$,
$\left(\frac{\hat{\mathbf{k}}}{M}-\frac{\hat{\mathbf{p}}}{m}\right)^{2}$,
$\left(\hat{\mathbf{r}}-\hat{\mathbf{q}}\right)\cdot\left(\frac{\hat{\mathbf{k}}}{M}-\frac{\hat{\mathbf{p}}}{m}\right)$,
$\left(\hat{\mathbf{r}}-\hat{\mathbf{q}}\right)\cdot\hat{\lambda}_{\mathbf{p}}$
and $\left(\frac{\hat{\mathbf{k}}}{M}-\frac{\hat{\mathbf{p}}}{m}\right)\cdot\hat{\lambda}_{\mathbf{p}}$
. 

Of the above, only $\left(\frac{\hat{\mathbf{k}}}{M}-\frac{\hat{\mathbf{p}}}{m}\right)^{2}$
commutes with the total linear momentum $\hat{\mathbf{k}}+\hat{\mathbf{p}}$.
Thus, in a Sudarshan hybrid only interactions that depend on the relative
canonical velocity conserve the total momentum. Moreover, only momentum
non-conserving interaction can have a quantum back reaction on the
classical observable variables since both $\hat{\mathbf{q}}$ and
$\hat{\mathbf{p}}$ commute with $\left(\frac{\hat{\mathbf{k}}}{M}-\frac{\hat{\mathbf{p}}}{m}\right)^{2}$.

\section{Discussion and Final Comments}

The present work does not tell whether a quantum-classical hybrid
theory is a good way to describe aspects of the real world or not.
What has been done here is to find what conditions a Sudarshan hybrid
must fulfill in order to be Galilei covariant. 

It is worth noting that in the literature there can be found interaction
terms that do not respect the commutation relations from the Galilei
algebra for the total system (see, for example, equation (13) in Ref
\cite{barcelo} or equation (18) in Ref \cite{peres}. See also \cite{last}).
It was noted in \cite{peres} that a quantum-classical coupling need
not to conserve the total energy, here a more general result was given
since conditions for the conservation of the total momentum were obtained.

The non-conservation of the total momentum for most of the allowed
interaction terms is a strange result. Galieli covariance guarantee
conservation of momentum in both quantum and classical mechanics but
the same Lie algebra has almost the opposite result for Sudarshan
hybrids. 

Restricting the theory so only interaction depending on the relative
velocity are considered is very harsh and seems to compromise any
actual application to real situations. On the other hand, allowing
position-dependent interactions requires the abandonment of a well
tested conservation law in both the classical and quantum regime.
It is true that there are new conserved quantities, like the total
translation operator $\hat{\mathscr{P}}=\hat{\mathbf{k}}+\hat{\lambda}_{\mathbf{q}}$.
However, the possible physical meaning of $\hat{\mathscr{P}}$ is
not clear as it depends on the classically unobservable $\hat{\lambda}_{\mathbf{q}}$.
The results derived in this work seem to put in serious doubt the
physical viability of Sudarshan hybrids as independent (non-quantum
derivable) theories.

\end{document}